\documentclass[twoside,twocolumn,9pt]{article}
\usepackage{extsizes}
\usepackage[super,sort&compress,comma]{natbib} 
\usepackage[version=3]{mhchem}
\usepackage[left=1.5cm, right=1.5cm, top=1.785cm, bottom=2.0cm]{geometry}
\usepackage{balance}
\usepackage{times,mathptmx}
\usepackage{sectsty}
\usepackage{graphicx} 
\usepackage{lastpage}
\usepackage[format=plain,justification=justified,singlelinecheck=false,font={stretch=1.125,small,sf},labelfont=bf,labelsep=space]{caption}
\usepackage{float}
\usepackage{fancyhdr}
\usepackage{fnpos}
\usepackage[english]{babel}
\addto{\captionsenglish}{%
}
\usepackage{array}
\usepackage{droidsans}
\usepackage{charter}
\usepackage[T1]{fontenc}
\usepackage[usenames,dvipsnames]{xcolor}
\usepackage{setspace}
\usepackage[compact]{titlesec}
\usepackage{hyperref}
\usepackage{amsmath,amssymb}
\usepackage{gensymb}

\usepackage{epstopdf}%This line makes .eps figures into .pdf - please comment out if not required.

\definecolor{cream}{RGB}{222,217,201}

\begin{document}

\pagestyle{fancy}
\thispagestyle{plain}

%%%PAGE SETUP - Please do not change any commands within this section%%%
\makeFNbottom
\makeatletter
\renewcommand\LARGE{\@setfontsize\LARGE{15pt}{17}}
\renewcommand\Large{\@setfontsize\Large{12pt}{14}}
\renewcommand\large{\@setfontsize\large{10pt}{12}}
\renewcommand\footnotesize{\@setfontsize\footnotesize{7pt}{10}}
\makeatother

\renewcommand{\thefootnote}{\fnsymbol{footnote}}
\renewcommand\footnoterule{\vspace*{1pt}% 
\color{cream}\hrule width 3.5in height 0.4pt \color{black}\vspace*{5pt}} 
\setcounter{secnumdepth}{5}

\makeatletter 
\renewcommand\@biblabel[1]{#1}            
\renewcommand\@makefntext[1]% 
{\noindent\makebox[0pt][r]{\@thefnmark\,}#1}
\makeatother 
\renewcommand{\figurename}{\small{Fig.}~}
\sectionfont{\sffamily\Large}
\subsectionfont{\normalsize}
\subsubsectionfont{\bf}
\setstretch{1.125} %In particular, please do not alter this line.
\setlength{\skip\footins}{0.8cm}
\setlength{\footnotesep}{0.25cm}
\setlength{\jot}{10pt}
\titlespacing*{\section}{0pt}{4pt}{4pt}
\titlespacing*{\subsection}{0pt}{15pt}{1pt}

\makeatletter 
\newlength{\figrulesep} 
\setlength{\figrulesep}{0.5\textfloatsep} 

\newcommand{\topfigrule}{\vspace*{-1pt}% 
\noindent{\color{cream}\rule[-\figrulesep]{\columnwidth}{1.5pt}} }

\newcommand{\botfigrule}{\vspace*{-2pt}% 
\noindent{\color{cream}\rule[\figrulesep]{\columnwidth}{1.5pt}} }

\newcommand{\dblfigrule}{\vspace*{-1pt}% 
\noindent{\color{cream}\rule[-\figrulesep]{\textwidth}{1.5pt}} }

\makeatother
%%%END OF FIGURE SETUP%%%

%%%%%%%%%%%%%%%%%%%%%%%%%%%%%%%%%%%%%%%%%%
% Alter some LaTeX defaults for better treatment of figures:
    % See p.105 of "TeX Unbound" for suggested values.
    % See pp. 199-200 of Lamport's "LaTeX" book for details.
    %   General parameters, for ALL pages:
    \renewcommand{\topfraction}{0.9}	% max fraction of floats at top
    \renewcommand{\bottomfraction}{0.8}	% max fraction of floats at bottom
    %   Parameters for TEXT pages (not float pages):
    \setcounter{topnumber}{2}
    \setcounter{bottomnumber}{2}
    \setcounter{totalnumber}{4}     % 2 may work better
    \setcounter{dbltopnumber}{2}    % for 2-column pages
    \renewcommand{\dbltopfraction}{0.9}	% fit big float above 2-col. text
    \renewcommand{\textfraction}{0.07}	% allow minimal text w. figs
    %   Parameters for FLOAT pages (not text pages):
    \renewcommand{\floatpagefraction}{0.7}	% require fuller float pages
	% N.B.: floatpagefraction MUST be less than topfraction !!
    \renewcommand{\dblfloatpagefraction}{0.7}	% require fuller float pages
    
    % remember to use [htp] or [htpb] for placement
%%%%%%%%%%%%%%%%%%%%%%%%%%%%%%%%%%%%%%%%%

%%%TITLE, AUTHORS AND ABSTRACT%%%
\twocolumn[
  \begin{@twocolumnfalse}
\vspace{3cm}
\sffamily
\begin{tabular}{m{1cm} p{14.5cm} }

& 
\noindent\LARGE{\textbf{Membrane-mediated interactions between arc-shaped particles strongly depend on membrane curvature}} %Article title goes here instead of the text "This is the title"
\\ \vspace{0.3cm} & \vspace{0.3cm} \\

 & \noindent\large{Francesco Bonazzi\textit{$^{a}$} and  Thomas R.\ Weikl\textit{$^{a}$}} \\

 & \\ 
 
  & \\ 
  
   & \\

& 
\noindent\normalsize{Besides direct molecular interactions, proteins and nanoparticles embedded in or adsorbed to membranes experience indirect interactions that are mediated by the membranes.
These membrane-mediated interactions arise from the membrane curvature induced by the particles and can lead to assemblies of particles that generate highly curved spherical or tubular membranes shapes, but have mainly been quantified for planar or weakly curved  membranes. In this article, we systematically investigate the membrane-mediated interactions of arc-shaped particles adsorbed to a variety of tubular and spherical membrane shapes with coarse-grained modelling and simulations. We determine both the pairwise interaction free energy, with includes entropic contributions due to rotational entropy loss at close particle distances, and the pairwise interaction energy without entropic components from particle distributions observed in the simulations. For membrane shapes with small curvature, the membrane-mediated interaction free energies of particle pairs exceed the thermal energy $k_BT$ and can lead to particle ordering and aggregation. The interactions strongly decrease with increasing curvature of the membrane shape and are minimal for tubular shapes with membrane curvatures close to the particle curvature.  
}

\\

 & \\ 

 & \\

\end{tabular}

 \end{@twocolumnfalse} \vspace{0.6cm}

  ]
%%%END OF TITLE, AUTHORS AND ABSTRACT%%%

%%%FONT SETUP - please do not change any commands within this section
\renewcommand*\rmdefault{bch}\normalfont\upshape
\rmfamily
\section*{}
\vspace{-1cm}

%%%FOOTNOTES%%%

\footnotetext{\textit{$^{a}$~Max Planck Institute of Colloids and Interfaces,  Am M\"uhlenberg 1, 14476 Potsdam Germany; email: thomas.weikl@mpikg.mpg.de}}

\section*{Introduction}

The intricately curved shapes of biological membranes are generated by protein assemblies \cite{Shibata09,Kozlov14,McMahon15,Baumgart11}. BAR (BIN/Amphiphysin/Rvs) domain proteins, for example, generate curvature by imposing their arc-like shape on membranes upon binding \cite{Peter04, Rao11, Mim12b, Simunovic19} and can induce membrane tubules \cite{Takei99} covered by dense protein coats \cite{Frost08,Mim12a,Adam15} or by less dense protein arrangements \cite{Daum16,LeRoux21}, depending on the protein type and concentration.  While the assembly of proteins or particles in solution  is typically driven by direct molecular interactions \cite{Shoemaker07,Marsh13,Min08,Grzelczak10}, assemblies of membrane-associated proteins or membrane-adsorbed particles  can also result from indirect interactions that are mediated by the membrane \cite{Weikl18,Johannes18,Haselwandter13,Idema19,Gao21}. 
Simulations and numerical approaches with a variety of different models \cite{Hafner19,Kumar22} indicate that such indirect interactions play a role in the tubulation of membranes by assemblies of arc-shaped proteins and particles  \cite{Simunovic13,Ramakrishnan13,Noguchi16,Noguchi17,Olinger16,Bonazzi19,Gao24}, 
can lead the cooperative wrapping of spherical  \cite{Bahrami12,Saric12b,Yue12,Raatz14,Raatz17,Xiong17,Tang18,Yan19,Zuraw19,Spangler21,Chen21} and elongated particles \cite{Raatz17,Xiong17},
and can result in assemblies of membrane-adsorbed Janus particles \cite{Reynwar07,Reynwar11,Bahrami18,Bahrami19,Zhu22,Zhu23,Sharma24}, elastic particles \cite{Midya23}, and hinge-like particles  \cite{Li22,Li24,Nambiar24}. Membrane-mediated interactions and assembly of spherical particles \cite{Koltover99,Wel16,Sarfati16,Wel17,Lavagna21,Azadbakht24b,Azadbakht24} as well as the cooperative wrapping of spherical virus-like particles \cite{Groza24} and rod-like particles \cite{Ham24} by membranes have also been observed in experiments. 
 
 The membrane-mediated interactions  result from a change of the membrane curvature induced by the particles or proteins and, therefore, can be expected to depend on the equilibrium curvature or shape of the membranes. However, membrane-mediated pair interactions have mainly been quantified for model proteins or particles that are adsorbed to or embedded in initially planar or weakly curved membranes \cite{Goulian93,Goulian93erratum,Park96,Netz97,Weikl98,Kim98,Dommersnes98,Dommersnes99a,Reynwar07,Reynwar11,Schweitzer15a,Yolcu12,Fournier15,Kahraman18}. Recent exceptions are the pair interactions of spherical Janus particles \cite{Bahrami18,Zhu22}, which have been found to depend on whether the particles adsorb to the inside or outside of the vesicles, and the pair interactions of spherical particles adsorbed to vesicles with different sizes \cite{Vahid17}.

In this article, we determine the membrane-mediated pair interactions of arc-shaped particles on tubular and spherical membrane vesicles from pair distributions of the particles observed in Monte Carlo simulations. In our coarse-grained model of membrane shaping, the membrane is described as a triangulated elastic surface, and the particles as segmented arcs that induce membrane curvature by binding to the membrane \cite{Bonazzi19}. In previous work, we found that the vesicle morphologies induced by the particles are determined by the arc angle and membrane coverage of the particles, and that membrane tubulation is induced by the particles above a threshold coverage of roughly 50\% \cite{Bonazzi19,Gao24}, in agreement with experimental observations for amphiphysin N-BAR domains \cite{LeRoux21}. In this work, in contrast, the vesicle morphology is fixed by the volume-to-area ratio of the vesicles, and the particle coverage in our simulations as adjusted to relatively small values between 5\% and 20\% to identify the particles' pair interactions. We determine the interaction free energy of particles bound to membrane tubules with different thicknesses and to spherical membrane vesicles by comparing the pair distributions of the particles obtained from simulations to ideal distributions of non-interacting particles. These interaction free energies include entropic components from the loss of rotational entropy of the arc-shaped particles at close distances, besides  the curvature-mediated interaction energies. In addition, we determine the interaction energies of the particles by comparing the pair distributions obtained from simulations to hard-cord distributions of `flattened' particles with the same shape, to quantify the curvature-mediated interaction of the particles without entropic components. We find that both the interaction free energies and the interaction energies strongly decrease with increasing curvature of the membrane shape and are minimal for tubular shapes with membrane curvatures close to the particle curvature.  

\section*{Results}

\begin{figure*}[h]
\centering
\includegraphics[width=\linewidth]{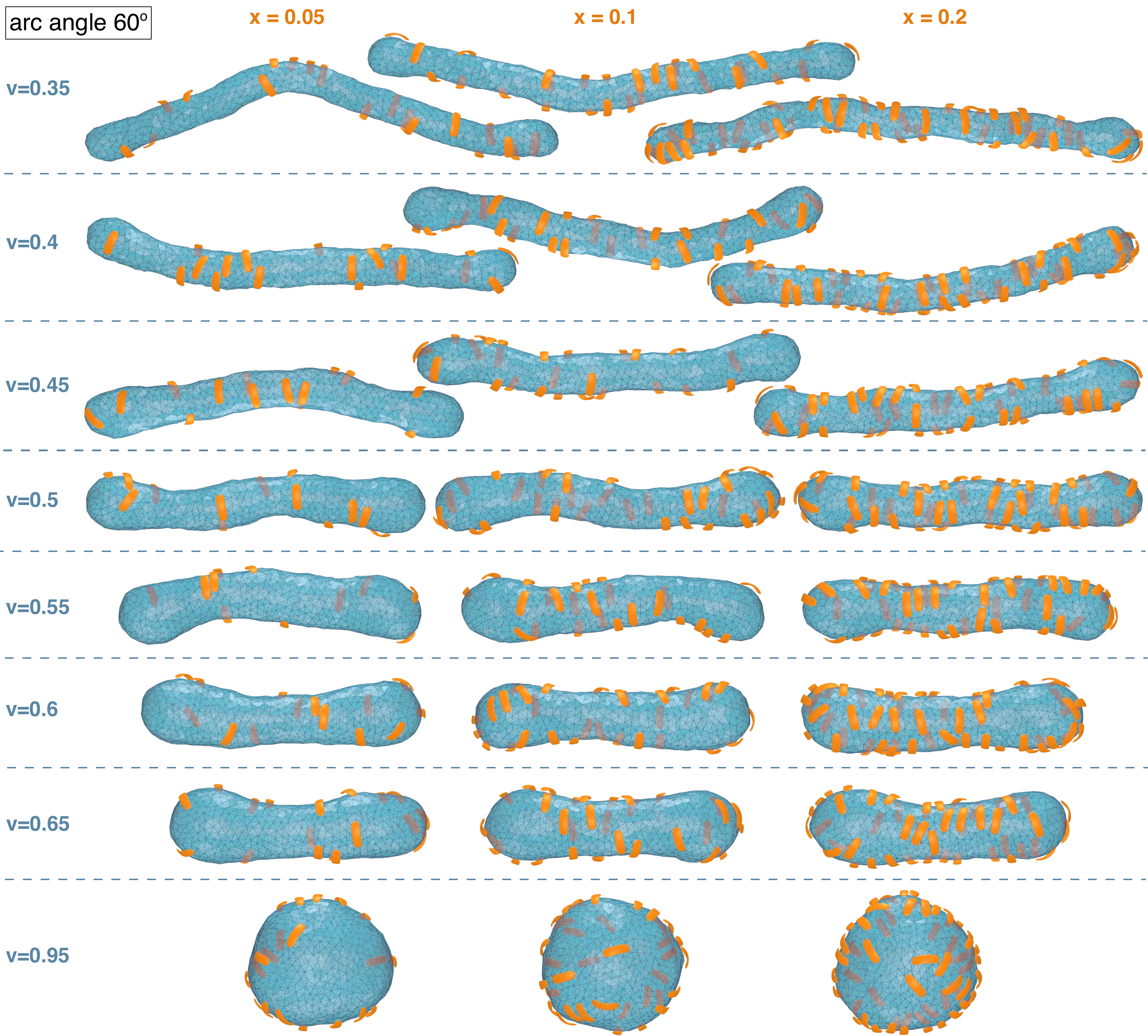}
  \caption{Exemplary simulation conformations for particles with arc angle $60\degree$ at different area coverages $x$ and reduced volume $v$ of the membrane.}
  \label{fig-conformations-60}
\end{figure*}

\begin{figure*}[h]
\centering
\includegraphics[width=\linewidth]{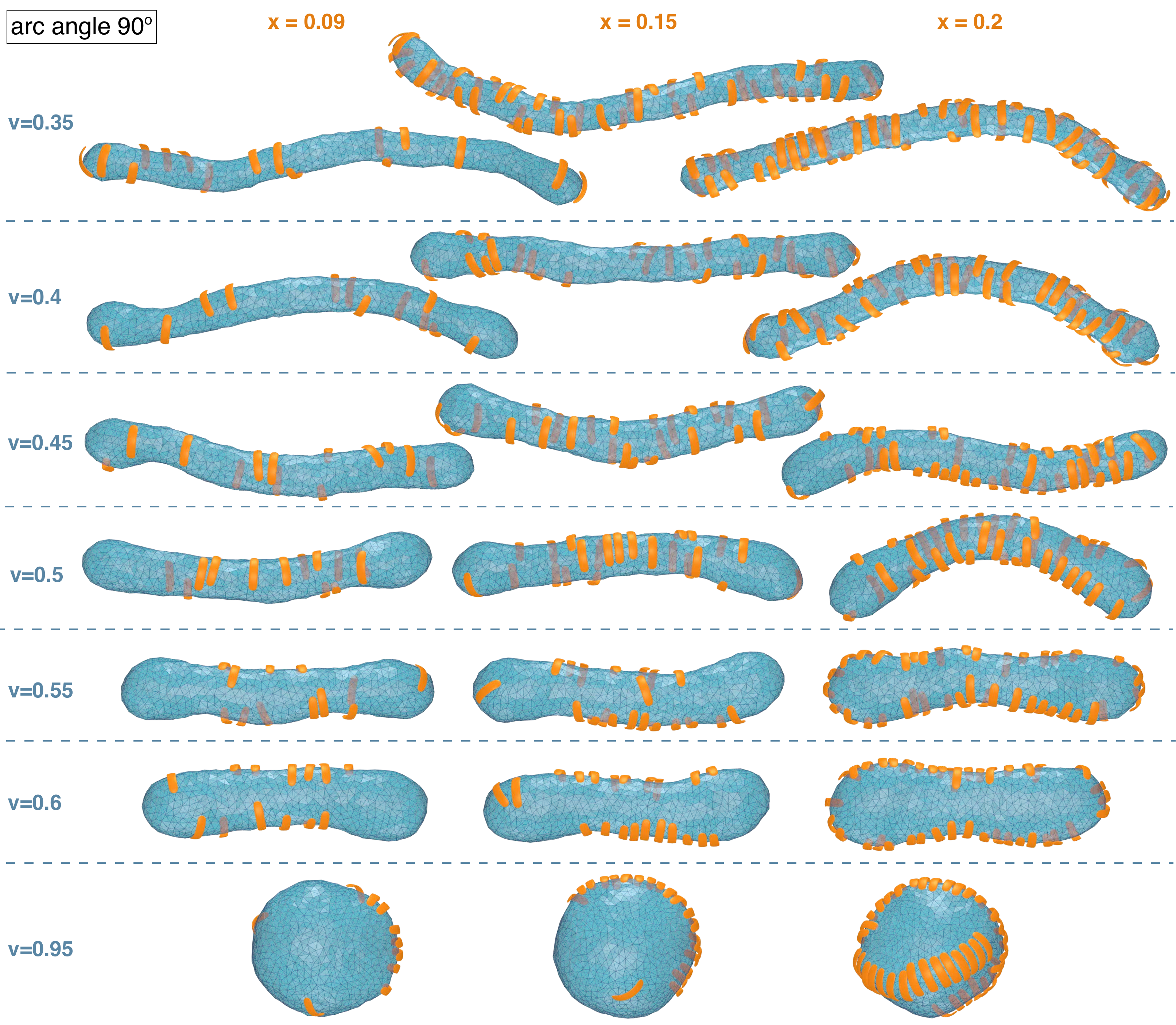}
  \caption{Exemplary simulation conformations for particles with arc angle $90\degree$ at different area coverages $x$ and reduced volume $v$ of the membrane.}
  \label{fig-conformations-90}
\end{figure*}

\subsection*{Conformations of arc-shaped particles on tubular and spherical vesicles}

Figs.\ \ref{fig-conformations-60} and  \ref{fig-conformations-90} illustrate simulations conformations of closed tubular and spherical membranes with bound arc-shaped particles. In our simulation model, the membranes are dynamically triangulated surfaces composed of 2000 triangles with a constrained total area $A$ (see Methods). The volume $V$ enclosed by the membrane is constrained to different values in the simulations, which results in the different membrane shapes  shown in Figs.\ \ref{fig-conformations-60} and  \ref{fig-conformations-90}. The fixed volume-to-area ratio of the membrane can be quantified  by the reduced volume $v = 6 \sqrt{\pi} V/A^{3/2}\le 1$, which adopts its maximum value of 1 for an ideal sphere. In our simulations of spherical membranes, we fix the reduced volume to $v = 0.95$ to allow for small variations and fluctuations around the overall spherical shape. In our simulations of spherocylindrical, tubular membranes, the reduced volume is fixed to $v=0.35$, $0.4$, $0.45$, $0.5$, $0.55$, $0.6$, or $0.65$. The tubular membrane shapes are metastable, because the stable membrane shape is an oblate ellipsoid for $0.592\lesssim v \lesssim 0.652$ and a stomatocyte for $0 < v \lesssim 0.592$ \cite{Seifert91,Bahrami17}. The metastable tubular shapes are protected against transformations into the stable membrane shapes by an energy barrier that depends on the bending rigidity $\kappa$ \cite{Bahrami17}. To ensure a sufficiently large energy barrier, we use the relatively large bending rigidity value $\kappa = 30\,k_B T$ from the range of typical bending rigidities of lipid membranes between about $10$ and $40\,k_BT$ \cite{Dimova14,Nagle13}.

The arc-shaped particles of our model are composed of either 3 or 4 quadratic segments with side length $a_p$ and an angle of $30\degree$ between adjacent segments. The arc angle of the particles, i.e.\ the angle between the terminal segments, thus is $60\degree$ for particles composed of 3 segments, and $90\degree$ for particles composed of 4 segments. The particles bind to the membrane with their inner, concave sides. A particle segment is bound to the triangulated membrane of our model if its distance to the closest membrane triangle is within a given range, and if the particle segment and membrane triangle are nearly parallel with an angle that is smaller than the cutoff angle $10\degree$ (see Methods for details). The relative area of the particle segments and membrane triangles is chosen such that a particle segment can only be bound to a single membrane triangle. The particles can bind to and unbind from the membranes in the simulation. In each simulation, the area fraction $x$ of the membrane covered by bound particles is kept at a constant value between 5\% and 20\% by dynamically adjusting the adhesion energy $U$ per particle segment. The total number of bound and unbound particles in our simulations is 400. 

The simulation conformations in Figs.\ \ref{fig-conformations-60} and  \ref{fig-conformations-90} illustrate a tendency of the particles to align side by side that increases with increasing reduced volume $v$ of the membrane and with increasing arc angle of the particles.  At the reduced volume $v = 0.95$,  particles with arc angle  90\degree form linear aggregates on the spherical membranes (see Fig.\ \ref{fig-conformations-90}). This alignment and aggregation of the particles is driven by membrane-mediated interactions, because the direct particle-particle interactions are purely repulsive in our model (see Methods).

\subsection*{Interaction free energies}

\begin{figure*}[h]
\centering
\includegraphics[width=\linewidth]{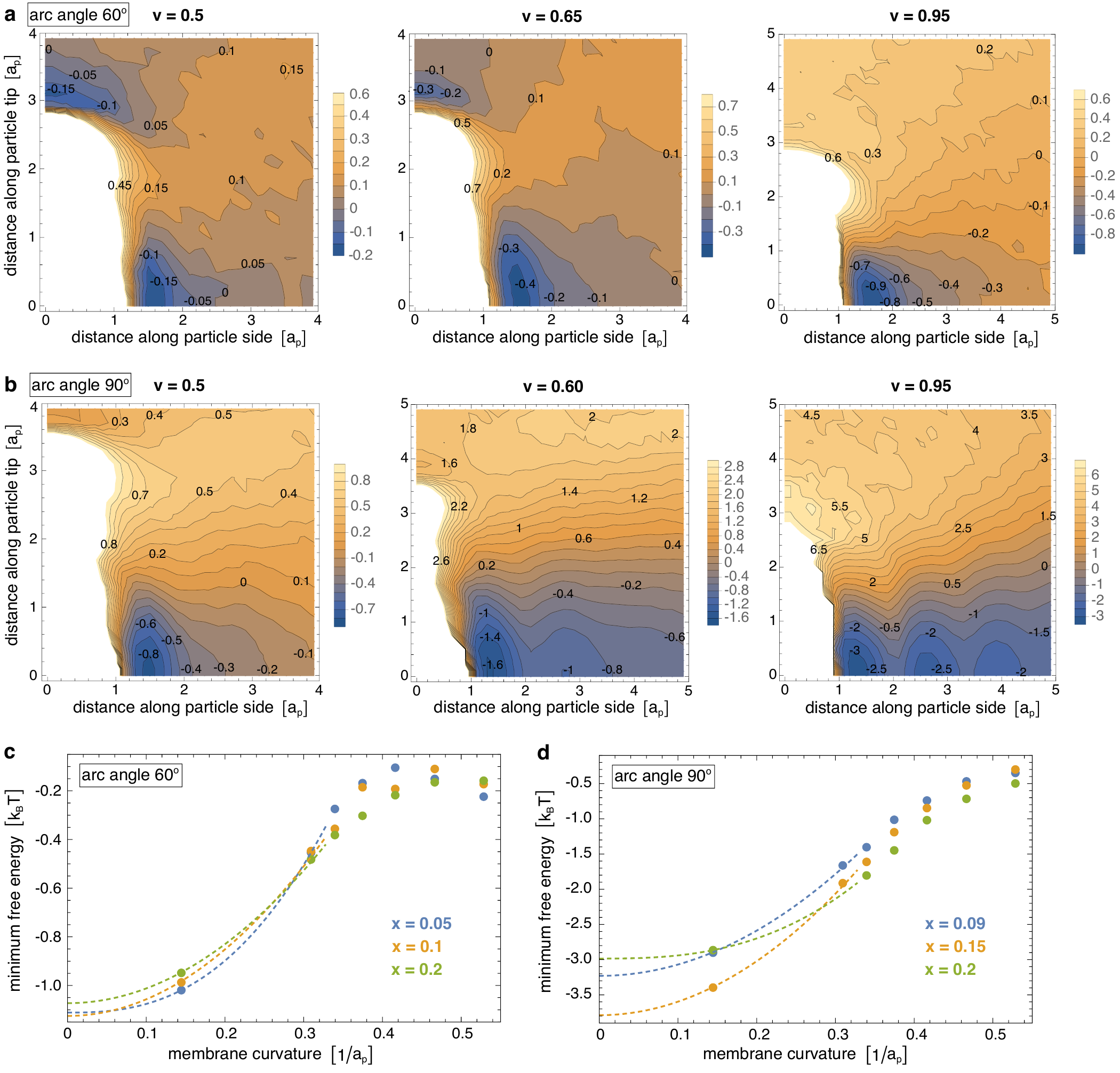}
  \caption{(a) Two-dimensional interaction free energy of pairs of arc-shaped particles with arc angle $60\degree$ for vesicle shapes with reduced volume $v = 0.5$, $0.65$, and $0.9$ and a particle coverage of 10\%. (b) Two-dimensional interaction free energy of pairs of arc-shaped particles with arc angle $90\degree$ for vesicle shapes with reduced volume $v = 0.5$, $0.6$, and $0.9$ and a particle coverage of 15\%.  (c,d)  Minimum interaction free energy {\it versus} membrane curvature for particles with arc angle  $60\degree$ and  $90\degree$ at different particle coverages of the membrane. 
For tubular membrane shapes, the membrane curvature here is calculated as the curvature for the tubular section of an ideal spherocylinder with the same area $A$ as the membrane in our simulations. For spherical shapes, the membrane curvature is calculated as the curvature of an ideal sphere with the same area $A$ as the membrane in our simulations. The dashed extrapolation lines to curvature 0 are guides for the eye.} 
  \label{fig-free-energy}
\end{figure*}

The membrane-mediated pair interactions of the particles can be quantified by two-dimensional interaction free energies, which we obtain from the two-dimensional pair distributions $P(x,y)$ of the particles observed in our simulations. These free energies reflect the highly anisotropic interactions of a particle along its side ($x$-direction) and along its tip ($y$-direction) to a neighboring particle, and are two-dimensional generalizations of the one-dimensional potential of mean force  \cite{Chandler87}. The two-dimensional interaction free energies are calculated as 
\begin{equation}
F(x,y) = -k_B T \ln\left[P(x,y)/P_{id}\right]
\end{equation}
where $P_{id}$ is the pair distribution of a non-interacting, ideal gas of particles (see Methods for details). These interaction free energies include entropic components from the reduction of the rotational entropy of the arc-shaped particles at close distances, besides  the curvature-mediated interaction energies. For tubular vesicle shapes with reduced volume $v \le 0.65$, we determine the pair distributions $P(x,y)$ in the central, tubular membrane segments and exclude particle pairs at the spherical ends (see Methods).

\begin{figure*}[t]
\centering
\includegraphics[width=\linewidth]{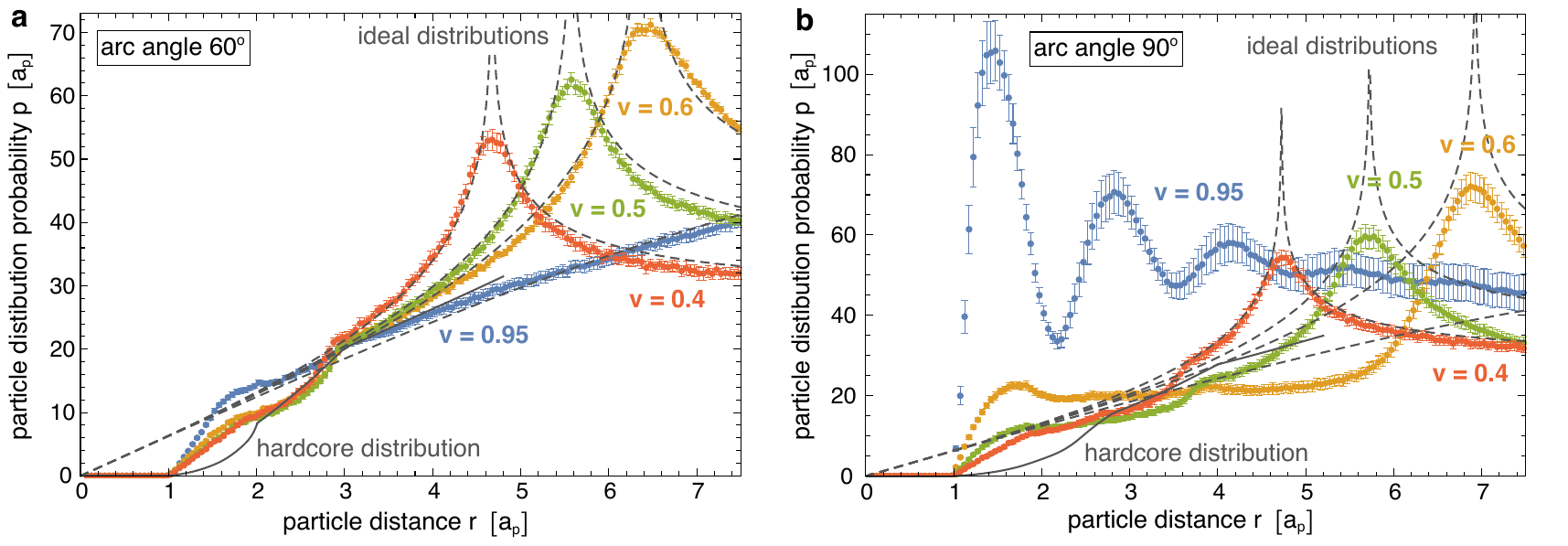}
  \caption{Radial pair distributions $p(r)$ of membrane-bound particles obtained from simulations with vesicle shapes for different reduced volumes $v$ (coloured data points),  ideal distributions of non-interacting point particles on the corresponding cylinders and spheres (grey dashed lines), and hardcore distribution $p_{hc}(r)$ of `flattened' particles with the same shape in a plane (full grey line). The area coverages of the membrane-bound particles are $x = 0.05$ in (a) for particles with arc angle $60\degree$ and $x = 0.09$ in (b) for for particles with arc angle $60\degree$. For the tubular membrane shapes with reduced volumes $v = 0.4$, $0.5$, and $0.6$, the pair distributions $p(r)$ exhibit maxima when the particle distance $r$ is equal to the diameter $2 R$. The radial pair distributions of ideal gas on a cylinder here are calculated as $p_{\rm ideal}(r) =  8 R\, f_{\rm elliptic}[0.5 \arccos[1 - r^2/(2 R^2)], 4 R^2/r^2]$ where $f_{\rm elliptic}$ is the elliptic integral of the first kind.}  
  \label{fig-distributions}
\end{figure*}

Fig.\ \ref{fig-free-energy}(a) and (b) display the resulting two-dimensional interaction free energies of the simulated particles with arc angle $60\degree$  and $90\degree$ at three different values of the reduced volume $v$ of the vesicles for intermediate particle coverages of 10\% and 15\%, respectively. At the small value $v= 0.5$, the vesicles adopt elongated, thin tubular shapes (see Figs.\ \ref{fig-conformations-60} and  \ref{fig-conformations-90}). At this value of $v$, the two-dimensional free energy of particles with arc angle $60\degree$ exhibits two minima with nearly equal depth of about $-0.15$ $k_BT$ for pair conformations in which the particles are oriented side-to-side and tip-to-tip, respectively (see Fig.\ \ref{fig-free-energy}(a)). At the intermediate value $v = 0.65$ at which the vesicles adopt thicker tubular shapes, the depths of the minima increases to about $-0.45$ $k_BT$ for  side-to-side alignment of particles with arc angle $60\degree$, and to about $-0.3$ $k_BT$ for tip-to-tip alignment. At $v = 0.95$, the vesicles adopt spherical shapes, and the depth of the minimum for side-to-side alignment of the particles further increases to nearly $-1.0$ $k_BT$, whereas tip-to-tip alignment of the particles is no longer energetically favourable. 

For particles with arc angle $90\degree$, the free-energy minima for pair conformations in which the particles are oriented side-to-side are clearly deeper compared to particles with arc angle $60\degree$, while free-energy minima for tip-to-tip alignment of the particles no longer occur (see Fig.\ \ref{fig-free-energy}(b)). These deeper free-energy minima for side-to-side orientation of the particles indicate clearly stronger curvature-mediated interactions of the particles, because the rotational entropy loss in side-to-side conformations is larger for the longer particles with arc angle $90\degree$, compared to particles with arc angle $60\degree$. At the reduced vesicle volume $v= 0.5$, the free-energy minimum for side-to-side alignment of the particles has a depth of about $-0.85$ $k_BT$. At the larger values $v= 0.6$ and $v= 0.95$, several minima in side-to-side direction of the particles appear, which reflects the tendency of the particles to form linear chains with side-to-side orientation in the simulations at these values of $v$ (see Fig.\ \ref{fig-conformations-90}).  The minimum at the smallest distance along the particle side reflects the interaction free energy of nearest neighbours and has a depth of about $-1.6$ $k_BT$ for $v= 0.6$ and $-3.4$ $k_BT$ for $v= 0.95$.

Fig.\ \ref{fig-free-energy}(c) and (d) illustrate how the interaction free-energy minima for side-to-side alignment of the particles depend on the curvature of the membrane at the three different particle coverages considered in our simulations. For tubular membrane shapes, the membrane curvature here is calculated as the curvature for the tubular section of an ideal spherocylinder with the same area $A$ as the membrane in our simulations. For spherical shapes, the membrane curvature is calculated as the curvature of an ideal sphere with the same area $A$ as the membrane in our simulations. For particles with arc angle $60\degree$, the interaction is weakest at membrane curvatures of about $0.45/a_p$, with a minimum interaction free energy of about $-0.15$ $k_B T$. These membrane curvature roughly corresponds to the induced curvature of the particles, which can be estimated from the equation $c = \theta/L$  for the curvature $c$ of a circular arc with angle $\theta$ and arc length $L$. For the average induced angle of $52.5\degree$ between membrane triangles bound to the terminal segments of a particle  \cite{Bonazzi19} and the particle arc length $L \simeq 2 a_p$ between the centers of these terminal segments with side length $a_p$, we obtain $c \simeq  0.46/a_p$. The interaction is strongest for spherical vesicles with a curvature of about $0.15/a_p$,  with a minimum interaction free energy of about $-1$ $k_B T$. The dashed lines in Fig.\ \ref{fig-free-energy}(c) are extrapolations to planar membranes with curvature $0$ and suggest pair interaction free-energy minima of about $-1.2$ $k_B T$ in planar membranes. 

For particles with arc angle  $90\degree$, the interaction is weakest at the largest membrane curvature of about $0.53/a_p$ obtained for the thinnest simulated tubules with reduced volume $v  = 0.35$ (see Fig.\ \ref{fig-free-energy}(d)). The interaction free-energy minima at this largest curvature range from about $-0.45$ to $-0.5$  $k_B T$, depending on the particle coverage $x$ of the membranes. For particles with  arc angle  $90\degree$, the induced curvature is about $c = \theta/L \simeq 0.48/a_p$ for an average induced angle $\theta \simeq 82.6\degree$ \cite{Bonazzi19} and arc length $L\simeq 3 a_p$ and, thus, slightly larger than for particles with  arc angle $60\degree$. The interaction is again strongest for spherical vesicles with a curvature of about $0.15/a_p$,  with a minimum interaction free energy in the range from  $-2.9$ to $-3.4$ $k_B T$.

\subsection*{Side-to-side interaction energy}

\begin{figure*}[h]
\centering
\includegraphics[width=\linewidth]{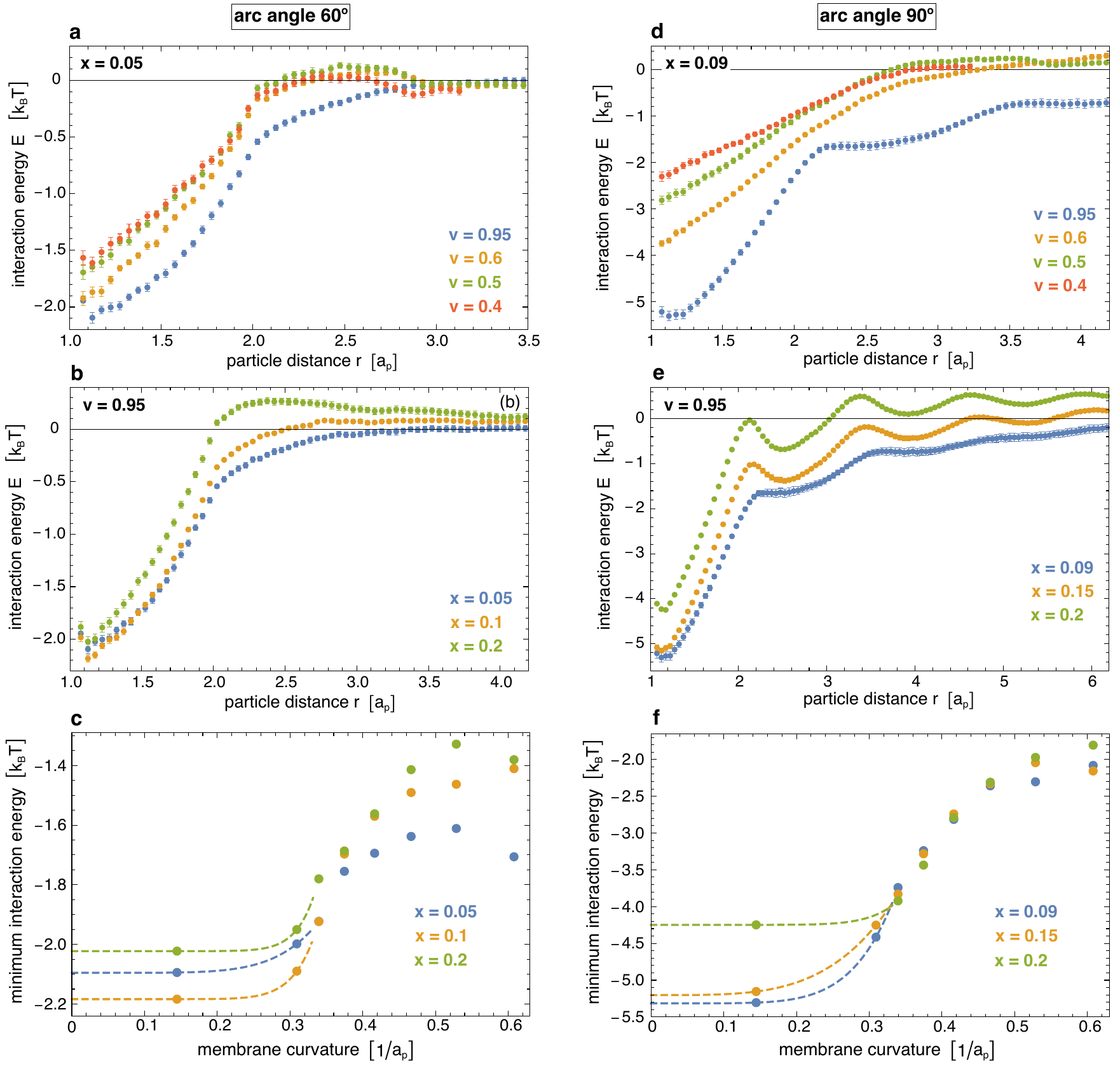}
  \caption{(a) and (b) Interaction energy profiles of pairs of arc-shaped particles with arc angle $60\degree$ at different values of the reduced volume $v$ of vesicles and of the coverage $x$ of the particles. (c) Minima of the interaction energy profiles {\it versus}  membrane curvature for particles with arc angle  $60\degree$ at coverages of $x =0.05$, $0.1$, and $0.2$. The membrane curvature here is calculated as in Fig.\  \ref{fig-free-energy}, and the dashed extrapolation lines to curvature 0 are guides for the eye.
 (d) and (e) Interaction energy profiles of pairs of arc-shaped particles with arc angle $90\degree$ at different values of $v$ and $x$. (f) Minima of the interaction energy profiles {\it versus}  membrane curvature for particles with arc angle  $90\degree$.
  }
  \label{fig-energies}
\end{figure*}

Besides interaction free energies of the particles, curvature-induced interaction energies without entropic components from rotational entropy losses can be obtained by comparing pair distributions $p$ from simulations to hard-cord distributions $p_{hc}$ of `flattened' particles with the same shape. We focus now on one-dimensional radial distributions and on the curvature-induced interaction energies for side-to-side aligned particle pairs, which are reflected by energetic minima at small separations that the particles can only achieve in side-to-side alignment. We calculate the one-dimensional, distance-dependent interaction energy of the particles as 
\begin{equation}
E(r) = -k_B T \ln\left[ \frac{p(r)}{p_{hc}(r)}\right]
\label{eq-E}
\end{equation}
with hard-core distributions $p_{hc}(r)$ determined for `flattened' particles with an angle of $0$ between the quadratic segment of side length $a_p$, rather than the angle of 30\degree\ of the curved, membrane-bound particles (see Methods for details). 

Fig.\ (\ref{fig-distributions}) illustrates the radial distributions $p(r)$ obtained from our simulations at different reduced vesicle volumes $v$ (coloured data points) as well as the hard-core distributions $p_{hc}(r)$ of the particles with arc angle  $60\degree$ and   $90\degree$ (full grey lines), and compares these distributions to the ideal distributions of non-interacting particles on cylinders for $v\le 0.6$ and spheres for $v\le 0.95$ (dashed grey lines). The distributions $p(r)$ and $p_{hc}$ are 0 at particle distances $r < a_p$ that are prevented by the hard-core interactions of the particles. At distances $r > a_p$, the hard-core distributions $p_{hc}(r)$ gradually increase and exhibit two kinks at distances $r$ at which the particles acquire more rotational freedom. For particles composed of 3 segments, the first kink of $p_{hc}(r)$ occurs at the distance $r = 2 a_p$ above which a particle that is located at the side of another particle gains full rotational freedom. The second kink occurs at the distance $r = 3 a_p$ above which a particle located at the tip of another particle has full rotational freedom. For particles composed of 4 segments, these kinks occur at corresponding, larger distances $r$.  The curvature-mediated interaction of the particles is reflected by clearly larger values of the distributions $p(r)$ at close distances $a_p < r < 3 a_p$ compared to the hard-core distributions $p_{hc}(r)$. For particles with arc angle $60\degree$, the radial distributions  $p(r)$ approach the ideal distributions at larger distances $r > 3 a_p$. For particles with arc angle $90\degree$, the tendency of the particles to form linear chains for reduced vesicle volume $v \gtrsim 0.5$ (see Fig.\ (\ref{fig-conformations-90})) leads to more global deviations from ideal distributions, and to multiple maxima  of $p(r)$  for spherical vesicles with $v = 0.95$.

Fig.\ (\ref{fig-energies}) illustrates the interaction energy profiles $E(r)$ and minimum interaction energies obtained from Eq.\ (\ref{eq-E}) for the distributions $p(r)$ obtained from our simulations and the numerically determined hard-core distributions $p_{hc}(r)$ at the corresponding particle coverages $x$ (see Methods for details). For particles with arc angle $60\degree$, the interaction energy profiles $E(r)$ tend to values close to 0 at larger distances $r > 3 a_p$ at which the distributions approach each other (see Figs.\ {\ref{fig-energies}(a,b) and \ref{fig-distributions}(a)). Our main aim are the minima of  $E(r)$ at short distances $r$ that reflect the curvature-mediated interaction energy of the particles in side-to-side alignment. These interaction energy minima strongly depend on the membrane curvature (see Fig.\ {\ref{fig-energies}(c)). As in Fig.\ \ref{fig-free-energy}(c), the interaction is weakest at membrane curvatures of about $0.45/a_p$ with minimum values from $-1.25$ to  $-1.6$ $k_B T$, depending on the particle coverage $x$. The interaction is strongest at the curvature of about $0.15/a_p$ of our spherical vesicles with minimum values from $-2.0$ to  $-2.2$ $k_B T$. The minimum values of the interaction energies are roughly $1$ $k_B T$ lower than the minima of the interaction free energies in Fig.\ (\ref{fig-free-energy}(c)) at corresponding curvatures. This difference of about $1$ $k_B T$ can be seen as the free-energy contribution from the loss of rotational entropy of the particles in the minimum-energy side-to-side pair conformations. 

For particles with arc angle $90\degree$, the minima of the interaction energy profiles $E(r)$ at short distances $r$ are clearly lower than for particles with arc angle $60\degree$  (see Figs.\ {\ref{fig-energies}(d) to (f)). For spherical vesicles with $v = 0.95$, the linear chains of particles observed in the simulations lead to multiple minima of $E(r)$ at the larger particle coverages $x = 0.15$ and $0.20$, and to more global deviations from the hard-core distributions. The global minimum values of the interaction energies are roughly $1$ to $2$ $k_B T$ lower than the minima of the interaction free energies in Fig.\ (\ref{fig-free-energy}(c)) at corresponding curvatures (see Figs.\ \ref{fig-free-energy}(d) and \ref{fig-energies}(f)), which reflects a slightly larger energy loss of rotational entropy for the longer particles with arc angle  $90\degree$, compared to particles with arc angle $60\degree$.

\section*{Discussion and Conclusions}

In this article, we have presented a general methodology to obtain the membrane-mediated interaction free energies and interaction energies of anisotropic curvature-inducing particles from distributions observed in simulations at relatively low membrane coverages of the particles. In this methodology, two-dimensional interaction free energies are calculated in comparison to ideal distributions of non-interacting particles, as generalization of the one-dimensional potential of mean force. The two-dimensionality of the interaction free energy is necessary to correctly capture the rotational entropy loss of the elongated particles at close contact, and the low coverages are required to identify the pair interactions of the particles, which should be independent of the membrane coverage of the particles. For the membrane coverages $x$ between $5$\% and  $20$\% in our simulations, the minima of the interaction free energy depend only rather weakly on $x$ (see Fig.\ \ref{fig-free-energy}(c) and (d)). Our main aim was to determine the dependence of the particles' interaction free energy  on the overall curvature to the membrane, which we adjusted by the reduced volume $v$ of membrane vesicles, from values of $v = 0.35$ for thin tubular vesicles to $v = 0.95$ for spherical vesicles (see Figs.\ \ref{fig-conformations-60} and \ref{fig-conformations-90}). We found that the pair interaction is smallest for tubular versicles with a curvature close to the particle curvature, with free energy minima of around $-0.2$ $k_B T$ and   $-0.5$ $k_B T$ for particles with arc angle $60\degree$ and  $90\degree$, respectively. The pair interaction is largest for  spherical vesicles with free energy minima of around  $-1$ $k_BT$ and  $-3$ $k_BT$ for the two types of particles.  We believe that the interaction free energies of our particles with arc angle $60\degree$ are realistic for BAR domain proteins such as the Arfaptin BAR domain and the Endophilin and Bin1 N-BAR domains, (i) because the induced average angle $52.5^\circ$ of these particles \cite{Bonazzi19} roughly corresponds to the angle enclosed by these BAR domain proteins \cite{Qualmann11}, and (ii) because the electron tomography images of membrane tubules induced by Bin1 N-BAR domains proteins show a rather loose protein arrangement with only short-ranged order, similar to the tubular morphologies induced by the particles in our simulations \cite{Bonazzi19,Gao24}. In general, the pair interaction free energy can be expected to depend also on the coupling between the particles and the membrane, which is determined by the particle-membrane interaction potential in our model (see Methods).  

Besides interaction free energies, we determined the interaction energies of the particles without entropic component from the distributions observed in our simulations and numerically determined hard-core distributions of `flattened' particles with the same shape. The minima of the interaction energy profiles are located at short particle distances at which the particles are aligned side-by-side and exhibit the same dependence on membrane curvature as the minima of the interaction free energies. However, the interaction energy minima are  about $-1$ $k_BT$ and between $-1$ to $-2$  $k_BT$ lower compared to the free energy minima obtained for the particles with arc angle $60\degree$ and $90\degree$, respectively. These differences between the minima of the interaction energies and interaction free energies result from the rotational entropy loss of the particles at close contact and can be seen to quantify the free energy contribution of these entropy loss.

\section*{Methods}
\subsection*{Model and simulations}

We model the membrane as a dynamically triangulated, closed surface. The membrane model is based on a standard discretization \cite{Julicher96,Bahrami12} of the bending energy ${\cal E}_\text{be} = 2\kappa \oint M^2 \,dS$ of closed fluid membranes\cite{Helfrich73} with bending rigidity $\kappa$ and local mean curvature $M$. Our discretized membranes are composed of $n_t = 2000$ triangles.  In our Monte Carlo (MC) simulations, the edges of the triangulated membrane are flipped to ensure membrane fluidity (dynamic triangulation), and the vertices of the triangulation are displaced to allow local changes of the membrane shape. The edge lengths of the triangles are kept within an interval $[a_m, \sqrt{3} a_m]$ to limit triangle distortions. The vertex displacements occur in MC steps in which a randomly selected vertex of the triangulated membrane is translated along a random direction by a distance that is randomly chosen from an interval between $0$ and $0.1 a_m$. We impose an overall tubular or spherical membrane shape by constraining both the membrane area $A$ and the enclosed volume $V$ of the membrane {\it via} harmonic constraining potentials. The volume-to-area ratio of the membrane therefore is fixed in our simulations. 

The discretized, arc-shaped particles of our model are linear chains of 3 or 4 identical planar quadratic segments with a side length $a_p$ and with an angle of $30\degree$ between neighboring segments that share a quadratic edge.\cite{Bonazzi19,Bonazzi21} The arc angle of the particles, i.e.\ the angle between the first and last segment, then adopts the values $60\degree$ and $90\degree$ for particles composed of 3 and 4 segments, respectively. Each planar segment of a particle interacts with the nearest triangle of the membrane {\em via} the particle-membrane adhesion potential \cite{Bonazzi19}
\begin{equation}
V_\text{pm} = - U f_r(r) f_\theta(\theta)
\label{Vpm}
\end{equation}
Here, $r$ is the distance between the center of the segment and the center of the nearest triangle, $\theta$ is the angle between the normals of the particle segment and this membrane triangle, and $U>0$ is the adhesion energy per particle segment. The distance-dependent function $f_r$ is a square-well function that adopts the values $f_r(r) = 1$ for $0.25\, a_m < r <  0.75\, a_m$ and $f_r(r)=0$ otherwise. The angle-dependent function $f_\theta$ is a square-well function with values $f_\theta(\theta) = 1$ for $|\theta| < 10\degree$ and $f_\theta(\theta) = 0$ otherwise. By convention, the normals of the membrane triangles are oriented outward from the enclosed volume of the membrane, and the normals of the particle segments are oriented away from the center of the particle arc. The particles thus bind with their inward curved, concave surface to the membrane. The overlapping of particles is prevented by a purely repulsive hard-core interaction that only allows distances between the centres of the planar segments of different particles that are larger than $a_p$. The hard-core area of a particle segment thus is $\pi a_p^2/4$. We choose the value $a_p = 1.5 a_m$ for the side length of the planar and quadratic particle segments. The particle segments then are slightly larger than the membrane triangles with minimum side length $a_m$, which ensures that different particle segments bind to different triangles. 

The positions and orientations of the particles are varied by particle translations and rotations in our simulations \cite{Bonazzi19}.  In a particle translation, a randomly selected particle is translated in random direction by a random distance between $0$ and $a_m$. In a particle rotation, a randomly selected particle is rotated around a rotation axis that passes through the central point along the particle arc. For particles that consist of 3 segments, the rotation axis runs through the center of the central segments. For particles composed of 4  segments, the rotation axis runs through the center of the edge that is shared by the two central segments. The rotation axis is oriented in a random direction. The random rotations are implemented using quaternions \cite{Frenkel02,Vesely82} with rotation angles between 0 and a maximum angle of about $2.3\degree$. Our simulations thus consist of  two types of MC steps for the membrane, vertex translations and edge flips, and two types of MC steps for translating and rotating the particles. The different types of MC steps occur with equal probabilities for single membrane vertices, edges, or particles \cite{Bonazzi19}. 

In each simulation, the area fraction $x$ of the membrane covered by bound particles is kept at a constant value between 5\% and 20\% by dynamically adjusting the adhesion energy $U$ per particle segment. The total number of particles in the system is N = 400. The area of the membrane is constrained to $A_0 \simeq 0.677 n_t a_m^2$. The strength of the harmonic constraining potential is chosen such that the fluctuations of the membrane area are limited to less than $1\%$. The simulations are run in a cubic box with periodic boundary conditions and volume $V_{\rm box} \simeq 3 \cdot 10^5 a_m^3$. The simulation box volume is 64 times as large as the volume of a perfect sphere with area $A_0$. After initial relaxation of the simulations until a steady state regarding the number of bound particles is reached, the simulation frames from which the particle distributions are determined are extracted from the simulations at time intervals of 100 Monte Carlo steps per vertex. For particles with arc angle $60\degree$, the total number of analysed conformations for tubular vesicles with reduced volume $v\le 0.65$ is about $180\, 000$ for the membrane coverage $x = 5\%$, $145\, 000$ for  $x = 10\%$, and $100\, 000$ for $x = 20\%$. For spherical vesicles with reduced volume $v = 0.95$, the total number of conformations is $150\, 000$, $85\, 000$, and $61\, 000$ at $x = 5\%$, $10\%$, and $20\%$, respectively. For particles with arc angle $90\degree$, the total number of analysed conformations for tubular vesicles with reduced volume $v\le 0.65$ is about $155\, 000$ for the membrane coverage $x = 9\%$, $125\, 000$ for  $x = 15\%$, and $90\, 000$ for $x = 20\%$. For spherical vesicles with reduced volume $v = 0.95$, the total number of conformations is $150\, 000$, $88\, 000$, and $59\, 000$ at $x = 9\%$, $15\%$, and $20\%$, respectively.

\subsection*{Interaction free energy}

We determined the two-dimensional interaction free energies of Fig.\ \ref{fig-free-energy} from two-dimensional pair distributions of the particles obtained from our simulations. These two-dimensional pair distributions reflect the distributions of neighbouring particles around a particle on the surface of the spherical or tubular membranes. For spherical membrane shapes, we determine the position of a neighbouring particle $j$ relative to a particle $i$ based on an angle $\phi$ between two planes P and Q. Plane P runs through the centres of the two terminal segments of particle $i$ and through the center of mass of the spherical membrane and, thus, reflects the orientation of particle $i$ on the membrane sphere. Plane Q runs through the centre of particle $i$, the center of particle $j$, and the center of mass of the spherical membrane.
For particles with arc angle 60\degree, which are composed of three segments, the particle center simply is the center of the central particle segment. For particles with arc angle 90\degree, which are composed of four segments, the particle center is the center of mass of the two central particle segments. The value $\phi = 0$ for the angle between planes P and Q indicates that the center of particle $j$ is located at the tip of particle $i$, and $\phi = \pi/2$ indicates that the center of particle $j$ is located at the side of particle $i$. The two-dimensional particle distributions then are calculated from the two coordinates $x = r \sin\phi$ and $y = r \cos\phi$ of all particle pairs where $r$ is the distance of the particle centres. To determine these distributions, we discretize the plane of the coordinates $x$ and $y$ into squares with side length $a_p/5$ and count the total number $M(x,y)$ of particle pairs with coordinate values in the square centered at $(x,y)$ for all conformations. The probability for finding a neighboring particle in the square centered at $(x,y)$ can then be determined as $P(x,y) = M(x,y)/(4 N_c N_b(N_b-1))$ where $N_c$ is the number of conformations and $N_b$ is the average number of bound particles in a conformation. The factor 4 takes into account that the angle $0\le \phi \le \pi/2$ is limited to one of the four angular quadrants because of the particle symmetry, and $N_b(N_b-1)$ is the average number of particle pairs per conformation that is considered in this calculation.
For ideal, non-interacting particles, the probability of finding a neighboring particle in square $(x,y)$ is $P_{id} = A_s/A$ where $A_s = (a_p/5)^2$ is the area of a discrete square and $A = A_0$ is the membrane area.
The two-dimensional interaction free energy then follows as 
\begin{equation}
F(x,y) = -k_B T \ln\left[ \frac{P(x,y)}{P_{id}}\right]  = -k_B T \ln\left[ \frac{M(x,y) A} {4 N_c  N_b(N_b - 1) A_s}\right]
\label{Fxy}
\end{equation}
This two-dimensional interaction free energy takes into account whether a neighboring particle is located at the side or at the tip of an arc-shaped particle, and is a generalization of the one-dimensional potential of mean force  \cite{Chandler87}.

For tubular membrane conformations, we define a central, tubular membrane segment of area $A= A_0/2$ based on the distance of the membrane triangles from the center of mass of the membrane, and focus on particle pairs $i$ and $j$ in which at least one particle is bound to the central membrane segment. Pairs of particles that are both bound to a spherical membrane end are, thus, excluded from the analysis. We next define the tubular axis of a conformation based on a singular value decomposition of the positions of all particles bound the membrane in this conformation, and project the particle centres on an ideal cylinder with radius $r_{cyl}$ around this tubular axis. For each particle pair, we determine the angle $\psi$ between the tubular axis and the vector that connects the projected particle centres, and define two coordinates as $x = r \cos\psi$ and $y =  r \sin\psi$  where $r$ is the distance of the (unprojected) particle centres. Because the particles are oriented on average perpendicular to the tubular axis, the value $\psi = 0$ corresponds to a side-to-side orientation of a particle pair, and the value $\psi = \pi/2$ corresponds to a tip-to-tip orientation. We discretize the plane of the two coordinates $x$ and $y$ as described above for spherical conformations, and determine the two-dimensional free energy $F(x,y)$ from Eq.\ (\ref{Fxy}) after rescaling the total number of particle pairs $M(x,y)$ by the factor $\sqrt{1 - y^2/(4 r_{cyl}^2)}$ to account for the cylindrical curvature in $y$-direction. For tubular conformations, $N_b$ in  Eq.\ (\ref{Fxy}) corresponds to the average number of particles bound to the central tubular membrane segments. In this calculation, the tubular radius $r_{cyl}$
of the particle layer around the membrane is obtained from the maximum of the one-dimensional distributions shown in Fig.\ \ref{fig-distributions}.

\subsection*{Interaction energy}

The one-dimensional interaction energies of Fig.\ \ref{fig-energies} were calculated based on Eq.\ (\ref{eq-E}). To determine the rescaled distribution $p(r)$ of the membrane-bound particles in this equation, we first determine the number of particle pairs  $M(r)$ with a distance of the particle centres within the shell $r \pm {\rm d}r/2$ with width ${\rm d}r = a_p/10$. The rescaled distribution then follows as  $p(r) = M(r)/({\rm d}r \,N_c N_b(N_b-1)/A)$ where $N_c$ is the number of simulation conformations and $N_b$ is the average number of bound particles in the considered membrane area $A$. For tubular membrane conformations, the area $A$ is the area  $A_0/2$ of the central tubular segment in which at least one of the bound particles in a pair needs to be located (see above). For spherical membrane conformations, the area $A$ is the total $A_0$ of the membrane.

\begin{figure}[h]
\centering
\includegraphics[width=\linewidth]{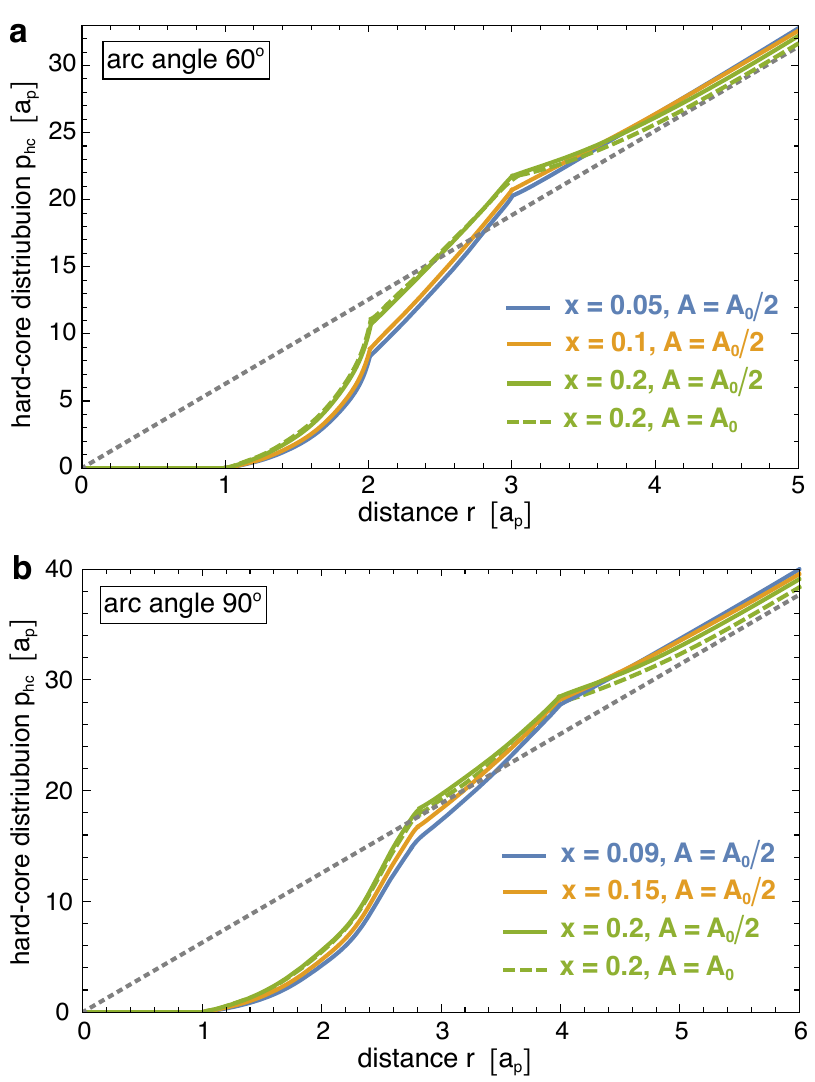}
  \caption{Hard-core distributions $p_{hc}(r)$ of particles with arc angle (a) $60\degree$ and (b) $90\degree$, numerically determined for corresponding `flattened particles' in a planar quadratic area $A$ at the indicated particle coverages $x$. For determining the interaction energy (\ref{eq-E}) of particles on spherical vesicles with reduced volume $v\le 0.95$, the area $A$ for calculating $p_{hc}(r)$ is taken to be the membrane area $A_0$, and the coverage $x$ is taken as the corresponding coverage of the simulations. In the case of tubular vesicles with  $v\le 0.65$, the area $A$ is the area $A_0/2$ of the central, tubular membrane segment used to calculate the distribution $p(r)$ from simulation conformations. The dashed lines represent the ideal distribution $p_{id}(r) = 2 \pi r$ of non-interacting  particles. }
  \label{fig-hc-distributions}
\end{figure}

The hard-core distributions $p_{hc}(r)$ in Eq.\ (\ref{eq-E}) are determined for `flattened' particles with an angle of $0$ between the quadratic segment of side length $a_p$, rather than the angle of $30\degree$ of the curved, membrane-bound particles. To obtain these distribution, we first generate a large number $N_c$ of non-overlapping conformations of $N_b$ `flattened' particles in a quadratic area $A$. Here, $N_b$ is the average number of membrane-bound particles and $A$ is the membrane in the corresponding membrane simulation system, in which $p(r)$ has been determined.  The hard-core distributions $p_{hc}(r)$ then are calculated from the number $M_{hc}(r)$ of particle pairs with a distance of the particle centres within the shell $r \pm {\rm d}r/2$ as $p_{hc}(r) = M_{hc}(r)/({\rm d}r \,N_c N_b(N_b-1)/A)$. We determine $M_{hc}(r)$ for standard periodic boundaries of the considered quadratic area $A$, i.e.\ we take the distance $r$ as the minimum distance among all periodic particle images. The hard-core distributions slightly depend on the area coverage $x = N_b A_p/A$ of the particles where $A_p = n_s a_p^2 \pi/4$ is the hard-core area of a single particle, which is composed of either $n_s = 3$ or $4$ segments. The hard-core distributions $p_{hc}(r)$ shown in Fig.\ \ref{fig-hc-distributions} slightly increase with area coverage $x$ at close distances 
$r < n_s a_p$ smaller than the particle length, because of an increase of the pressure in the particle gas with $x$. This pressure pushes the particles together, against their entropic repulsion at distances $r < n_s a_p$.  Because of the normalization implied by the definition of the distributions, higher values of $p_{hc}(r)$ at small distances $r < n_s a_p$ lead to lower values at large distances. The hard-core distributions exhibit two kinks at distances $r$ at which the particles acquire more rotational freedom. For particles composed of 3 segments, the first kink of $p_{hc}(r)$ occurs at the distance $r = 2 a_p$ above which a particle that is located at the side of another particle gains full rotational freedom. The second kink occurs at the distance $r = 3 a_p$ above which a particle located at the tip of another particle has full rotational freedom. For particles composed of 4 segments, these kinks occur at corresponding, larger distances $r$.  

\section*{Conflicts of interest}
There are no conflicts to declare.

\section*{Acknowledgements}
Financial support from the Max Planck Society and from the Deutsche Forschungsgemeinschaft (DFG) {\em via} the International Research Training Group 1524 ``Self-Assembled Soft Matter Nano-Structures at Interfaces" is gratefully acknowledged.

%%%END OF MAIN TEXT%%%

%The \balance command can be used to balance the columns on the final page if desired. It should be placed anywhere within the first column of the last page.

\balance

\clearpage

%If notes are included in your references you can change the title from 'References' to 'Notes and references' using the following command:
%\renewcommand\refname{Notes and references}

%%%REFERENCES%%%

\providecommand*{\mcitethebibliography}{\thebibliography}
\csname @ifundefined\endcsname{endmcitethebibliography}
{\let\endmcitethebibliography\endthebibliography}{}

\end{document}